\documentclass[pra,a4paper,aps,twocolumn,showpacs,superscriptaddress,groupedaddress]{revtex4}
\usepackage[dvips]{graphicx}
\usepackage{ae}
\usepackage[T1]{fontenc}
\usepackage[ansinew]{inputenc}
\usepackage{amsmath}
\usepackage{amssymb}
\usepackage{graphicx}
\usepackage{capt-of}
\usepackage{color}
\usepackage{adjustbox}
\usepackage[colorlinks=true,linkcolor=red,citecolor=blue]{hyperref}
\usepackage{lscape}
\begin{document}
	\title{Genuine Einstein-Podolsky-Rosen steering of three-qubit states by multiple sequential observers}
	
\author{Shashank Gupta}
\email{shashankg687@bose.res.in}
\affiliation{S. N. Bose National Centre for Basic Sciences, Block JD, Sector III, Salt Lake, Kolkata 700 106, India}

\author{Ananda G. Maity}
\email{anandamaity289@gmail.com}
\affiliation{S. N. Bose National Centre for Basic Sciences, Block JD, Sector III, Salt Lake, Kolkata 700 106, India}

\author{Debarshi Das}
\email{dasdebarshi90@gmail.com}
\affiliation{S. N. Bose National Centre for Basic Sciences, Block JD, Sector III, Salt Lake, Kolkata 700 106, India}

\author{Arup Roy}
\email{arup145.roy@gmail.com}
\affiliation{S. N. Bose National Centre for Basic Sciences, Block JD, Sector III, Salt Lake, Kolkata 700 106, India}

\author{A. S. Majumdar}
\email{archan@bose.res.in}
\affiliation{S. N. Bose National Centre for Basic Sciences, Block JD, Sector III, Salt Lake, Kolkata 700 106, India}

\date{\today}

\begin{abstract}
We investigate the possibility of multiple use of a single copy of three-qubit state for detecting genuine tripartite Einstein-Podolsky-Rosen (EPR) steering. A pure three-qubit state of either the Greenberger-Horne-Zeilinger (GHZ)-type or W-type is shared between two fixed observers in two wings and a sequence of multiple observers in the third wing who perform unsharp or non-projective measurements. The measurement settings of each of the multiple observers in the third wing is independent and uncorrelated with the measurement settings and outcomes of the previous observers. In such set-up, we investigate all possible types of ($2\rightarrow1$) and ($1\rightarrow2$) genuine tripartite EPR steering. For each case, we obtain an upper limit on the number of observers on the third wing who can demonstrate genuine EPR steering through the quantum violation of an appropriate tripartite steering inequality. We show that the GHZ state allows for a higher number of observers compared to that for W states.  Additionally, ($1\rightarrow 2$) genuine steering is possible for a larger range of the sharpness parameters compared to that for the ($2 \rightarrow 1$) genuine steering cases. 

\end{abstract}

\maketitle

\section{INTRODUCTION}

Quantum network, as its name suggests, is composed of multiple observers, and is the future of secure quantum communication  tasks where multipartite quantum correlation serves as the resource. Utilization and characterization of such resources stemming from multipartite entanglement \cite{KarolZ} is rather important for  information theoretic applications \cite{Mur,Hil,Sor,Rau,Sca,Zha,ap3,Bri,Hyl,toth2} as well as from foundational perspectives. A multipartite state is called genuinely entangled \cite{Guh} if and only if  it cannot be written as a convex linear combination of states, each of which is separable  with respect to some partition. However, characterizing entanglement across each partition becomes more complex when the number of parties is increased.

Well known methods of characterizing genuine multipartite entanglement include genuine multipartite entanglement witnesses \cite{Acin,Bruss,Bou,Guh,entropic}. Some
of these methods though require certain prior information about the state and, more importantly, the underlying assumption of trusted preparation and measurements devices  \cite{Rosset}. Alternatively, the approach of genuine multipartite entanglement detection based on the violations of multipartite Bell-type inequalities \cite{svet,See,Coll,Uff,Sev,Nag,Pal,Ban2,Ban1,Lia} can be adopted since genuine multipartite entanglement is necessary (but not sufficient) for genuine multipartite Bell-nonlocality. However, detection of genuine multipartite Bell-nonlocality requires high detector efficiencies and low levels of noise. Hence, in practical scenarios, it may not always be feasible to implement such strategies for utilizing genuine multipartite entanglement. 

A substitute method to expose genuine multipartite entanglement relies upon the concept of Einstein-Podolsky-Rosen (EPR) steering \cite{steerreview}. The idea of EPR steering was first introduced in the bipartite scenario by Schr$\ddot{o}$dinger \cite{Schr1,Schr2} in the context of the  EPR argument \cite{epr}, where the choice of measurement settings  on one side can `steer' the state on the other side. Much later, Reid proposed a criterion for experimentally demonstrating the EPR argument using the Heisenberg uncertainty relation \cite{Reid}. Subsequently, Wiseman \textit{et al.} \cite{Wise1,Wise2} presented an information theoretic perspective  of EPR steering. Several criteria have since been  proposed in order to detect bipartite EPR steering \cite{Cavalcanti1,Cavalcanti2,Pramanik1,Pramanik,Pramanik2,Cavalcanti3,Maity1,DDJ+17}. Demonstrating quantum steering is crucial  for various applications in information processing and communication tasks \cite{branciard,app1,app3,app4,app5,app6,supic,jeba,peng}.
 
Recently, the concept of genuine EPR steering has gained attention in multipartite scenarios \cite{stm1,Li,Daniel,stm2,stm3,stm4}. Detection of genuine multipartite EPR steering certifies the presence of genuine multipartite entanglement since genuine multipartite entanglement is necessary (but not sufficient) for genuine multipartite EPR steering.  This idea of utilizing  genuine multipartite EPR steering for detecting genuine entanglement, being an intermediate concept between genuine multipartite entanglement witnesses and genuine multipartite Bell nonlocality based approaches, is a lot more noise resistant than the genuine entanglement detection by the violations Bell-type inequalities and is less experimentally demanding than the standard genuine entanglement witness methods as precise control over the measurement devices is not required for all observers in case of EPR steering.

Genuine multipartite EPR steering acts as the primary resource in hybrid quantum networks where some observers have more trust on their measuring devices than the others. These types of hybrid quantum networks are the most natural building blocks for practical and commercial quantum information processing tasks where the general consumers  may not want to trust their providers, for example, in the context of  quantum internet \cite{Kimble2008}, multipartite secret sharing \cite{Armstrong2015,sss1,sss2}, commercial quantum key distribution and commercial random number generation \cite{Daniel}.  Genuinely multipartite EPR steerable states are thus of  importance  for future quantum technologies.

Due to the difficulties in quantum state preparations and the ubiquitous decoherence effect, preserving genuine multipartite EPR steerability in a state even after performing local quantum operations is a major obstacle in  quantum information protocols. In this regard,  the advantages of nondestructive sequential quantum measurements in quantum communication schemes cannot be overstated \cite{Bergou2013}. In the present context, the challenges in preparing multipartite steerable states \cite{Armstrong2015,Mattar2017,Liu2020}  make it natural to ask whether one copy of genuinely EPR steerable state can be sequentially used multiple times when some quantum advantage is gained in each round. Specifically, our goal in the present study is to address whether genuine multipartite EPR steering can be detected by multiple observers who can sequentially access a single copy of a multipartite entangled state. In other words, we are motivated by the question as to whether it is possible to genuinely steer a single copy of a resourceful $m$-party ($m > 2$) state multiple times sequentially by $n$ observers ($n > m$).

The question of sharing of quantum correlations by multiple sequential observers was first posed by Silva \textit{et al.} \cite{sygp} in the context of bipartite Bell nonlocality. The scenario contains  an entangled pair of two spin-$\frac{1}{2}$ particles, shared between two spatially separated wings. Alice performs projective measurement on one half of the entangled state and multiple Bobs perform unsharp measurements on the other half sequentially and independently of each other. Considering unbiased frequencies of the inputs of the each Bob, it was conjectured \cite{sygp} that at most two Bobs can demonstrate Bell-nonlocality with a single Alice. This result was subsequently confirmed  \cite{majumdar}  using the unsharp measurement formalism \cite{pb1,pb2}.  Note that it is possible to increase further  the number of Bobs by choosing different sharpness parameters for the two inputs of each Bob, as shown recently in \cite{colbeck}.  

The approach enunciated  in the  work of Silva \textit{et al.} \cite{sygp} based on unbiased frequencies of inputs for each Bob,
has been realized experimentally \cite{exp1,exp2}. Subsequently, this idea of sharing quantum correlations by multiple sequential observers has been applied to  EPR steering \cite{sas, shenoy, Choi2020}, entanglement detection \cite{bera,Foletto,malnew}, steerability of local quantum coherence \cite{saunak},  violations of various Bell-type inequalities \cite{das,rennew,expnew}, preparation contextuality \cite{pcon,akpan},  unbounded randomness generation \cite{ran}, distinguishing quantum theory from classical simulations \cite{cc}, quantum  teleportation \cite{sroy}, and random access codes \cite{rac,exp4,rac1}.

The above studies dealing with the issue of sharing quantum correlations are restricted to two spatially separated particles. Quantum correlations possess special features for tripartite systems, due to their monogamous character which is not prevalent in bipartite states \cite{Dey13,Jeba18,Gupta18}. Recently, the possibility of sequential detection of genuine tripartite Bell nonlocality by multiple observers has been studied by Saha \textit{et al.} \cite{Saha}. On the other hand, Maity \textit{et al.} \cite{Maity2} have addressed sequential detection of genuine tripartite entanglement by multiple observers using genuine tripartite entanglement witnesses \cite{Acin,Bruss,Guh} for Greenberger-Horne-Zeilinger (GHZ) state \cite{GHZ} and W state \cite{wstate}, as well as by the violation of Bell type inequalities.  However, implementation of genuine tripartite EPR steering by multiple observers has hitherto remained an open question, which is of further interest because of the inherent asymmetry or directionality of EPR steering. 

 Here, we should note that sequential detection of genuine tripartite Bell nonlocality \cite{Saha} implies sequential detection of genuine tripartite EPR steering. However, genuine EPR steering being a weaker correlation than genuine Bell nonlocality, it is important to investigate whether genuine EPR steering can be detected by a larger number of sequential parties compared to that in case of genuine Bell nonlocality. On the other hand, sequential detection of genuine entanglement \cite{Maity2} does not always imply sequential detection of genuine steering as genuine steering is not necessary for demonstrating genuine entanglement. These issues motivate the present study. 

The scenario investigated in the present work consists of three spin-$\frac{1}{2}$ particles, spatially separated and shared between three wings. In the third wing there exist a sequence of observers  who perform non-projective measurements. The choice of measurement settings for each observer in the third wing is independent and uncorrelated with the measurement settings and outcomes of the previous observers. We consider that the initially shared three-qubit state is either of the GHZ type \cite{GHZ} or of W type \cite{wstate}. We consider all possible types of ($2\rightarrow1$) and ($1\rightarrow2$) genuine tripartite steering in the above set-up, investigating the maximum number  of  observers on the third wing for whom it is possible to demonstrate genuine tripartite EPR steering through the violations of the appropriate inequalities proposed by Cavalcanti \textit{et al.}  \cite{Daniel}. For each case, we obtain an upper limit on the number of observers on the third wing. We find out that this bound is greater using GHZ state compared to that for W state. Moreover, it turns out that the range of values of the sharpness parameters enabling genuine steering by the multiple observers turns out to be greater for the ($1 \rightarrow 2$) steering cases  compared to that for the ($2 \rightarrow 1$) steering cases.

The plan of the paper is as follows: in Section \ref{s2} we present the basic tools for detecting genuine tripartite EPR steering. The measurement framework involving multiple sequential observers used in this paper is also described in this Section. In Section \ref{s3}, we present the main analysis of this paper, and discuss the results obtained in the context of sequential detection of genuine tripartite EPR steering of the initially shared three-qubit GHZ state as well as the W state in all possible cases of tripartite steering scenario. Finally, we conclude in Section \ref{s4}.

\section{Preliminaries} \label{s2}

In this section we discuss in brief the concept of genuine tripartite EPR steering and  the measurements employed in order to probe sequential implementation of genuine tripartite steering by multiple observers.

\subsection{Detection of Genuine Tripartite Steering}
 Let us consider that a tripartite state $\rho$ is shared among three observers, say, Alice, Bob and Charlie. In this scenario either Alice tries to genuinely steer Bob's and Charlie's particles ($1 \rightarrow 2$ steering) or Alice-Bob try to genuinely steer Charlie's particle ($2 \rightarrow 1$ steering).

In the first case, i.e., when Alice tries to genuinely steer Bob-Charlie,  Alice's measurement operators are denoted by $A_{a|x}$, where $x$ is the choice of input and $a$ is the outcome. After Alice's measurement, each element of the set $\{\sigma_{a|x}^{BC}\}_{a,x}$ of unnormalized conditional states (assemblage)  on Bob-Charlie's end is given by,
 \begin{equation}
	\sigma_{a|x}^{BC}= \text{tr}_A\big[(A_{a|x} \otimes \mathbb{I}_B \otimes \mathbb{I}_C) \, \rho \big].
\label{assemblage1}
\end{equation}

Now, if the initial state $\rho$ is not genuinely entangled, then it is in the following bi-separable form,
\begin{align}
	\rho &= \sum_{\lambda}p_{\lambda}^{A:BC}\rho_{\lambda}^A \otimes \rho_{\lambda}^{BC} + \sum_{\mu}p_{\mu}^{B:AC}\rho_{\mu}^B \otimes \rho_{\mu}^{AC} \nonumber \\
	& \quad + \sum_{\nu}p_{\nu}^{AB:C}\rho_{\nu}^{AB} \otimes \rho_{\nu}^{C}.
	\label{biseparableform}
\end{align}
Here $p_{\lambda}^{A:BC}$, $p_{\mu}^{B:AC}$ and $p_{\nu}^{AB:C}$ are probability distributions where $A:BC$, $B:AC$ and $AB:C$ 
symbolise the different types of bipartition's. For example, $A:BC$ represents the bipartition between Alice and Bob-Charlie. Similarly, $AB:C$ represents the bipartition between Alice-Bob and Charlie. Here, no distinction is made between $AB:C$ and $C:AB$.
  
When the initial state $\rho$ is not genuinely entangled, then  each element (\ref{assemblage1}) of the assemblage $\{\sigma_{a|x}^{BC}\}_{a,x}$ is of the following form,
\begin{align}
	\sigma_{a|x}^{BC} &= \text{tr}_A \big[ (A_{a|x} \otimes \mathbb{I}_B \otimes \mathbb{I}_C) \, \rho \big] \nonumber \\
				&= \sum_{\lambda} p_{\lambda}^{A:BC}  p_{\lambda}(a|x) \rho_{\lambda}^{BC} \nonumber \\
				& \quad + \sum_{\mu}p_{\mu}^{B:AC}\rho_{\mu}^B \otimes \sigma_{a|x\mu}^C \nonumber \\
				& \quad + \sum_{\nu}p_{\nu}^{AB:C} \sigma_{a|x\nu}^B \otimes \rho_{\nu}^{C} \quad \forall \, a,x. 
				\label{ass1}
\end{align}
Here $p_{\lambda}(a|x)$ denotes the probability of getting the outcome $a$ when Alice performs the measurement denoted by $x$ on the state $\rho_{\lambda}^A$; $\sigma_{a|x\mu}^C$ denotes the unnormalized conditional state on Charlie's side when Alice gets the outcome $a$ by performing the measurement denoted by $x$ on the bipartite state $\rho_{\mu}^{AC}$ shared between Alice and Charlie; and $\sigma_{a|x\nu}^B$ denotes the unnormalized conditional state on Bob's side when Alice gets the outcome $a$ by performing the measurement denoted by $x$ on the bipartite state $\rho_{\nu}^{AB}$ shared between Alice and Bob. 

   Note that each element of the  ensemble $\{\sigma_{a|x}^{BC}\}_{a,x}$ given by Eq.(\ref{ass1}) is
expressed as a sum of three terms. The first term implies that there is
no steering from Alice to Bob-Charlie. The other two terms consist of separable states of Bob and Charlie. In each of these two terms,
Alice can steer either Bob's subsystem or Charlie's subsystem, but not both the subsystems. 
This can be viewed in terms of  a hybrid-local-hidden-state model. In the first term, the hidden variable $\lambda$ predetermines the global state of Bob and
Charlie, i.e., $\rho_{\lambda}^{BC}$, and this state may be
entangled. In the second term, the hidden variable $\mu$ predetermines the state of Bob, but not the state of Charlie. In the last term, the state of Charlie, but not the state of Bob, is predetermined by the hidden variable $\nu$.

 If any element of the assemblage $\{\sigma_{a|x}^{BC}\}_{a,x}$ cannot be written in the form (\ref{ass1}), then the assemblage indicates that the state $\rho$ possesses genuine tripartite entanglement.  In Refs. \cite{Daniel,stm2,stm3,stm4}, this has been taken as the definition of genuine tripartite EPR steering. In other words, when an element  of the assemblage $\{\sigma_{a|x}^{BC}\}_{a,x}$ cannot be written in the form (\ref{ass1}), then the assemblage demonstrate genuine  tripartite EPR steering from Alice to Bob-Charlie. This definition is motivated from the fact that genuine tripartite EPR steering from Alice to Bob-Charlie should be considered as the certification of genuine tripartite entanglement when Alice's measuring devices are uncharacterized. This is similar to the bipartite case, where bipartite steering from Alice to Bob is considered as the certification of entanglement when Alice performs uncharacterized measurements.

 Next, consider the second case where Alice-Bob try to genuinely steer Charlie, Alice's  measurement operators are denoted by $A_{a|x}$ and Bob's measurement operators are denoted by $B_{b|y}$. Here, $x$ and $y$ are the choices of inputs by Alice and Bob respectively, and $a$ and $b$ are the outcomes of Alice's and Bob's measurements respectively. After their measurements,  each element of the set $\{\sigma_{ab|xy}^{C}\}_{a,b,x,y}$ of unnormalized conditional states (assemblage) prepared on Charlie's side is given by,
 \begin{equation}
	\sigma_{ab|xy}^{C}= \text{tr}_{AB} \big[(A_{a|x} \otimes B_{b|y} \otimes \mathbb{I}_C) \, \rho \big].
\label{assemblage2}
\end{equation}

In this case, when the initial state $\rho$ is not genuinely entangled, i.e., when $\rho$ can be written in the form (\ref{biseparableform}),  then each element of the assemblage $\{\sigma_{ab|xy}^{C}\}_{a,b,x,y}$  can be written as
\begin{align}
	\sigma_{ab|xy}^{C} &= \text{tr}_{AB} \big[(A_{a|x} \otimes B_{b|y} \otimes \mathbb{I}_C) \, \rho \big] \nonumber \\
				&= \sum_{\lambda} p_{\lambda}^{A:BC} p_{\lambda}(a|x) \sigma_{b|y\lambda}^{C} \nonumber \\
				& \quad + \sum_{\mu}p_{\mu}^{B:AC} p_{\mu}(b|y)  \sigma_{a|x\mu}^C \nonumber \\
				&\quad + \sum_{\nu}p_{\nu}^{AB:C} p_{\nu}(ab|xy) \rho_{\nu}^{C} \quad \forall \, a,b,x,y. 
				\label{ass2}
\end{align}
Here $p_{\lambda}(a|x)$ denotes the probability of getting the outcome $a$ when Alice performs the measurement denoted by $x$ on the state $\rho_{\lambda}^A$; $\sigma_{b|y\lambda}^{C}$ denotes the unnormalized conditional state on Charlie's side when Bob gets the outcome $b$ by performing the measurement denoted by $y$ on the bipartite state $\rho_{\lambda}^{BC}$ shared between Bob and Charlie; $p_{\mu}(b|y)$ denotes the probability of getting the outcome $b$ when Bob performs the measurement denoted by $y$ on the state $\rho_{\mu}^B$; $\sigma_{a|x\mu}^C$  denotes the unnormalized conditional state on Charlie's side when Alice gets the outcome $a$ by performing the measurement denoted by $x$ on the bipartite state $\rho_{\mu}^{AC}$ shared between Alice and Charlie; $p_{\nu}(ab|xy)$ denotes the joint probability of getting the outcomes $a$ and $b$ when Alice and Bob perform the measurements denoted by $x$ and $y$, respectively, on the bipartite state $\rho_{\nu}^{AB}$ shared between Alice and Bob. 

 In this case, each element of the  assemblage $\{\sigma_{ab|xy}^{C}\}_{a,b,x,y}$  is expressed as a convex sum of three terms. In the first term, only Bob can steer Charlie, but Alice cannot steer Charlie. In the second term, only Alice can steer the state of Charlie, but Bob cannot steer Charlie's state. Finally, in the third term, Alice and Bob cannot jointly steer Charlie's state, but they can share quantum correlations between themselves. 

If any element of the assemblage $\{\sigma_{ab|xy}^{C}\}_{a,b,x,y}$ cannot be written in the form (\ref{ass2}), then the assemblage indicates  the presence of genuine tripartite entanglement in the initial state $\rho$.  The definition of genuine tripartite EPR steering from Alice-Bob to Charlie can be presented based on the decomposition (\ref{ass2}). When an element  of the assemblage $\{\sigma_{ab|xy}^{C}\}_{a,b,x,y}$ cannot be written in the form (\ref{ass2}), then the assemblage demonstrates genuine  tripartite EPR steering from Alice-Bob to Charlie \cite{Daniel,stm2,stm3,stm4}.

With the motivation of obtaining experimentally testable genuine tripartite steering witnesses based on the above analysis, Cavalcanti \textit{et al.} \cite{Daniel} designed several inequalities  which detect genuine steering of GHZ and W states in the two cases mentioned above. These  are  genuine EPR steering inequalities. These inequalities detect the genuine steering solely from the measurement correlations in the steering scenario, thus acting as experimentally testable genuine steering witnesses. Characterization of assemblage is not required which is resource-intensive and experimentally challenging. So, we have taken these inequalities as tools to detect genuine steering from measurement correlations only at each sequential step. 

If the shared state $\rho$ is three-qubit GHZ state, then genuine tripartite EPR steering from Alice to Bob-Charlie is detected by the quantum violation of the following inequality,
\begin{align}
G_1 &= 1 + g_{\alpha} \langle Z_BZ_C\rangle - \frac{1}{3} ( \langle A_3Z_B \rangle + \langle A_3Z_C \rangle  \nonumber \\
& + \langle A_1X_BX_C\rangle - \langle A_1Y_BY_C\rangle - \langle A_2X_BY_C\rangle \nonumber \\
& - \langle A_2Y_BX_C\rangle )\geq 0
\label{GHZ1}
\end{align}
with $g_{\alpha}=0.1547$, and  $A_i$ for $i=1, 2, 3$, being the observables associated with Alice's measurements with outcomes $\pm 1$ and $X$, $Y$ and $Z$ represent Pauli operators. The GHZ state violates the inequality by $-0.845$ when
Alice's measurements are $X$, $Y$ and $Z$, which numerical optimization suggests are the optimal choices for Alice. Though this inequality is the most suitable for GHZ state, its quantum violation by any state implies  genuine tripartite EPR steering from Alice to Bob-Charlie.

 If the shared state $\rho$ is a three-qubit GHZ state, then one can use the following inequality to detect genuine tripartite EPR steering from Alice-Bob to Charlie,
\begin{align}
G_2 &= 1 - \alpha (\langle A_3B_3\rangle +  \langle A_3Z \rangle + \langle B_3Z \rangle) - \beta( \langle A_1B_1X \rangle \nonumber \\
& -  \langle A_1B_2Y\rangle -  \langle A_2B_1Y\rangle -   \langle A_2B_2X\rangle )\geq 0
\label{GHZ2}
\end{align}
where $\alpha = 0.183$, $\beta = 0.258$, $B_i$ for $i=1, 2, 3$ represents the observables associated with Bob's measurements with outcomes $\pm 1$. The GHZ state violates the above inequality by $-0.582$ when Alice and Bob both perform $X$, $Y$ and $Z$ measurements.  The above inequality is satisfied by all assemblages having decomposition (\ref{ass2}) and its quantum violation implies genuine tripartite EPR steering from Alice-Bob to Charlie.

 On the other hand, if the shared state $\rho$ is a three-qubit W state,  then the suitable inequality for demonstrating genuine tripartite EPR steering from Alice to Bob-Charlie is given by,
\begin{align}
W_1 &= 1 + w_{\alpha}(\langle Z_B \rangle + \langle Z_C \rangle) - w_{\beta}  \langle Z_BZ_C\rangle   \nonumber \\
& - w_{\gamma} ( \langle X_B X_C \rangle + \langle Y_B Y_C \rangle + \langle A_3 X_B X_C \rangle + \langle A_3 Y_B Y_C \rangle )  \nonumber \\
& + w_{\delta} ( \langle A_3 \rangle +   \langle A_3Z_BZ_C \rangle ) + w_{\epsilon} ( \langle A_3Z_B\rangle + \langle A_3Z_C \rangle )  \nonumber \\
& - w_{\phi}( \langle A_1X_B\rangle + \langle A_1X_C\rangle + \langle A_2Y_B\rangle + \langle A_2Y_C\rangle  \nonumber \\ 
& + \langle A_1X_BZ_C\rangle + \langle A_1Z_BX_C\rangle ) + \langle A_2Y_BZ_C\rangle + \langle A_2Z_BY_C\rangle ) \nonumber \\
& \geq 0
\label{W1}
\end{align}
where $w_{\alpha}= 0.4405$, $w_{\beta} = 0.0037$,  $w_{\gamma} = 0.1570$, $w_{\delta}= 0.2424$, $w_{\epsilon} = 0.1848$, $w_{\phi}=0.2533$, 
with the pure W state achieving the violation $-0.759$.  The above inequality is satisfied by all assemblages having decomposition given by Eq.(\ref{ass1}) and its quantum violation implies genuine tripartite EPR steering from Alice to Bob-Charlie.

Similarly, the suitable inequality for demonstrating genuine tripartite EPR steering from Alice-Bob to Charlie in case of three-qubit W state is given by,
\begin{align}
W_2 &= 1 + w_{\kappa}(\langle A_3 \rangle + \langle B_3 \rangle) + w_{\lambda} \langle Z\rangle - w_{\eta}( \langle A_1 X \rangle \nonumber \\
&+ \langle A_2 Y \rangle + \langle B_1 X  \rangle + \langle B_2 Y  \rangle ) + w_{\mu} ( \langle A_3 Z \rangle + \langle B_3 Z \rangle) \nonumber \\
& - w_{\nu} ( \langle A_1B_1\rangle + \langle A_2B_2 \rangle ) + w_{\omega}  \langle A_3B_3\rangle \nonumber \\
& - w_{\pi}( \langle A_1B_1Z\rangle + \langle A_2B_2Z\rangle ) + w_{\theta} \langle A_3B_3Z\rangle  \nonumber \\ 
& -w_{\xi} (\langle A_1B_3X\rangle + \langle A_2B_3Y\rangle + \langle A_3B_1X\rangle + \langle A_3B_2 Y\rangle ) \nonumber \\
& \geq 0
\label{W2}
\end{align}
where, $w_{\kappa}=0.2517$, $w_{\lambda}=0.3520$, $w_{\eta}=0.1112$, $w_{\mu}=0.1296$,
$w_{\nu}=0.1943$, $w_{\omega}=0.2277$, $w_{\pi}=0.1590$, $w_{\theta}=0.2228$, $w_{\xi}=0.2298$,
with the pure W state achieving the violation $-0.480$.  Violation of this inequality implies genuine tripartite EPR steering from Alice-Bob to Charlie.

\subsection{Sequential measurement context} \label{scenario}
We now describe further the measurement context adopted throughout the present paper for sequential detection of genuine tripartite EPR steering. Let us consider  a tripartite system of state $\rho$ (either GHZ state or W state) consisting of spatially separated three spin-$\frac{1}{2}$ particles.  Specifically, we consider the following two scenarios:

{\bf Scenario A- Multiple Alices performing sequential measurements:} In this case, we consider multiple Alices (Alice$^1$, Alice$^2$, $\cdots$, Alice$^n$) perform measurements on the first particle sequentially. On the other hand, a single Bob and a single Charlie  perform projective measurements on the second and third particle, respectively. Here we ask the following two questions:

1) How may Alices can genuinely steer Bob-Charlie?

2) How many Alices, together with the single Bob, can genuinely steer Charlie?

Since our aim is to explore how many Alices can demonstrate genuine tripartite  steering through the violation of genuine EPR steering inequalities (\ref{GHZ1}, \ref{GHZ2}, \ref{W1}, \ref{W2}), multiple Alices cannot perform projective measurements. If any Alice performs a projective measurement, then the genuine EPR steerability of the state will be completely lost and the next Alice cannot demonstrate genuine EPR steering. However, no such restriction is required for the measurements performed by the last Alice in the sequence. Hence, for $n$ number of Alices, the first $(n - 1)$ Alices in the sequence should perform weak measurements.  Note that the no-signalling condition (the probability of obtaining one party's outcome does not depend on the other spatially separated party's setting) is satisfied between any Alice$^m$ ($m \in \{1, 2, \cdots, n \}$), Bob and Charlie. However, this condition is not satisfied between multiple Alices. In fact, Alice$^1$  implicitly signals to Alice$^2$  by her  choice of measurement on the state before she  passes it on and, similarly, Alice$^2$ signals to Alice$^3$, and so on.

{\bf Scenario B- Multiple Charlies performing sequential measurements:} In this case, we consider multiple Charlies (Charlie$^1$, Charlie$^2$, $\cdots$, Charlie$^n$) perform measurements on the third particle sequentially. On the other hand, a single Alice and a single Bob perform projective measurements on the first and second particle respectively.  In this case, we ask the following two questions:

1) How many Charlies, along with the single Bob, can be  genuinely steered by Alice?

2) How many Charlies  can be  genuinely steered by Alice-Bob?

We will address these questions using the genuine EPR steering inequalities (\ref{GHZ1}, \ref{GHZ2}, \ref{W1}, \ref{W2}). Here the measurements performed by Charlie$^1$, Charlie$^2$, $\cdots$, Charlie$^{n-1}$ are unsharp and the measurement performed by Charlie$^n$ is projective.  Here the no-signalling condition  is satisfied between Alice, Bob and Charlie$^m$ ($m \in \{1, 2, \cdots, n \}$). However, this condition is not satisfied for multiple Charlies.

In each of the above cases, we further make the following two assumptions.
First, each of the multiple observers (either multiple Alices or multiple Charlies)  performs measurement on the same particle independently of other prior observers. In other words, Alice$^m$ (Charlie$^m$) with $m$ $\in$ $\{1, 2, \cdots, n\}$ is ignorant of the choices of measurement settings and outcomes of Alice$^1$, Alice$^2$, $\cdots$, Alice$^{m-1}$ (Charlie$^1$, Charlie$^2$, $\cdots$, Charlie$^{m-1}$). Secondly, we restrict ourselves to the unbiased input scenario which implies that  all possible measurement settings of each of the multiple observers (either multiple Alices or multiple Charlies) are equally probable. 
 Note that we have not considered multiple Bobs as the role of Bob is equivalent to that of either Alice or Charlie in each of the above sequential steering scenarios.

\begin{figure}[t!]
\centering
\includegraphics[scale=0.4]{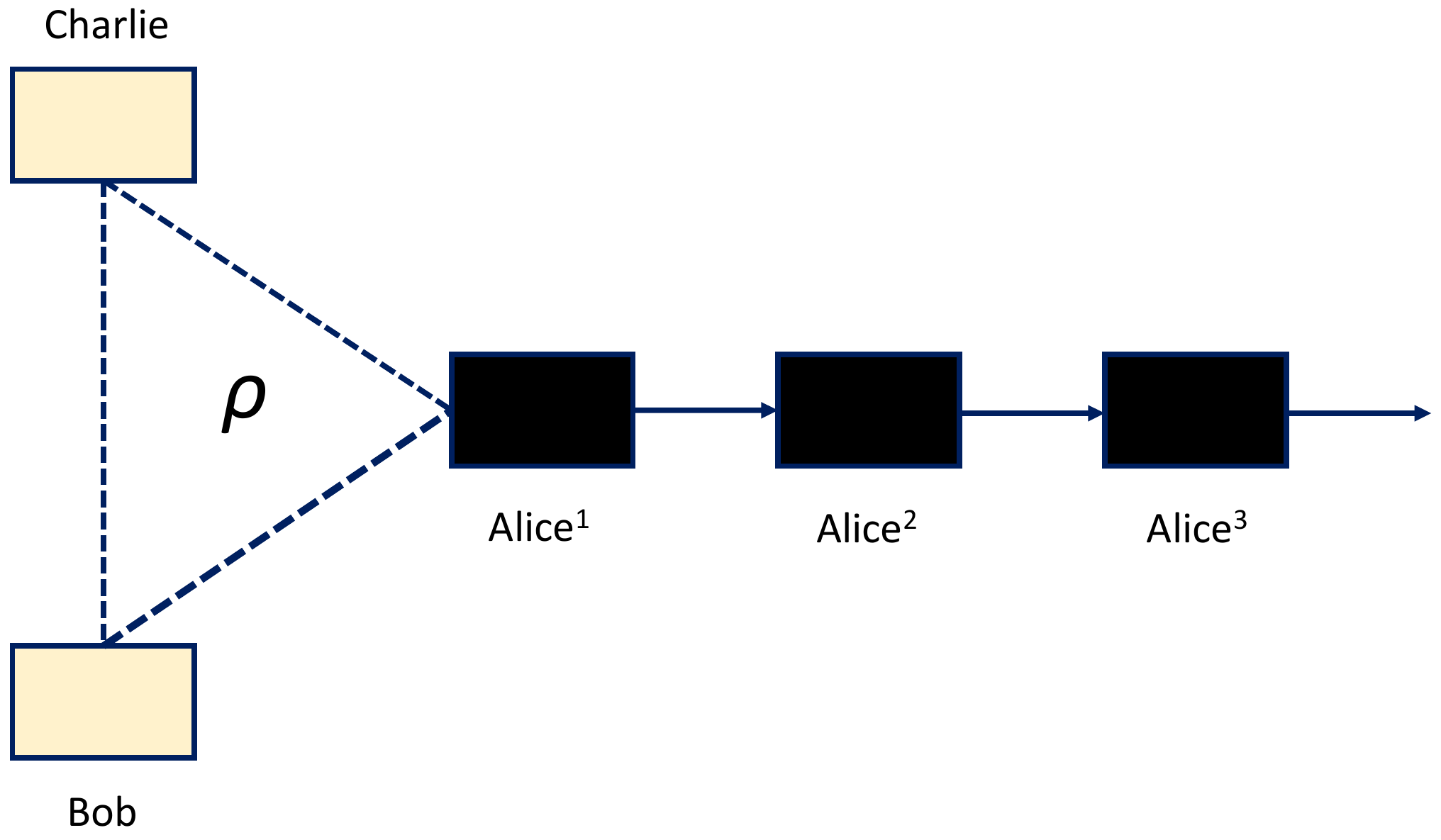}
\caption{(Color Online) Sequential detection of genuine tripartite EPR steering with multiple Alices in the scenario where Alice tries to demonstrate genuine tripartite ($1 \rightarrow 2$) steering to Bob-Charlie. Three spatially separated spin-$\frac{1}{2}$ particles,  prepared in the three-qubit state $\rho$,  are shared between multiple Alices (Alice$^1$, Alice$^2$, $\cdots$, Alice$^n$), a single Bob and a single Charlie. }
\label{fig1}
\end{figure}

\begin{figure}[t!]
\centering
\includegraphics[scale=0.4]{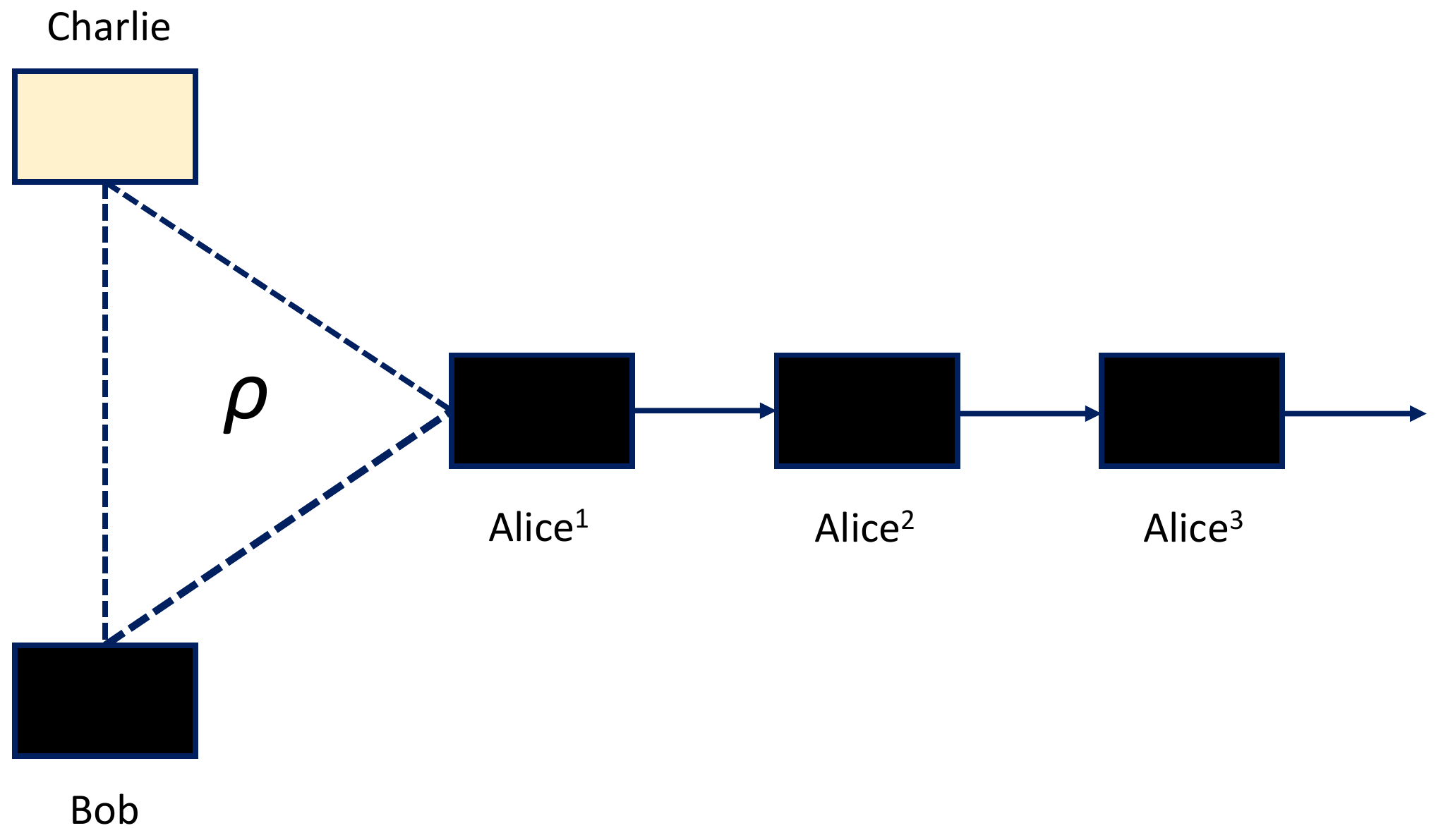}
\caption{(Color Online) Sequential detection of genuine tripartite EPR steering with multiple Alices in the scenario where both Alice and Bob try to demonstrate genuine tripartite ($2 \rightarrow 1$) steering to Charlie. Three spatially separated spin-$\frac{1}{2}$ particles, prepared in the three-qubit state $\rho$,  are shared between multiple Alices (Alice$^1$, Alice$^2$, $\cdots$, Alice$^n$), a single Bob and a single Charlie. }
\label{fig2}
\end{figure}

\begin{figure}[t!]
\centering
\includegraphics[scale=0.4]{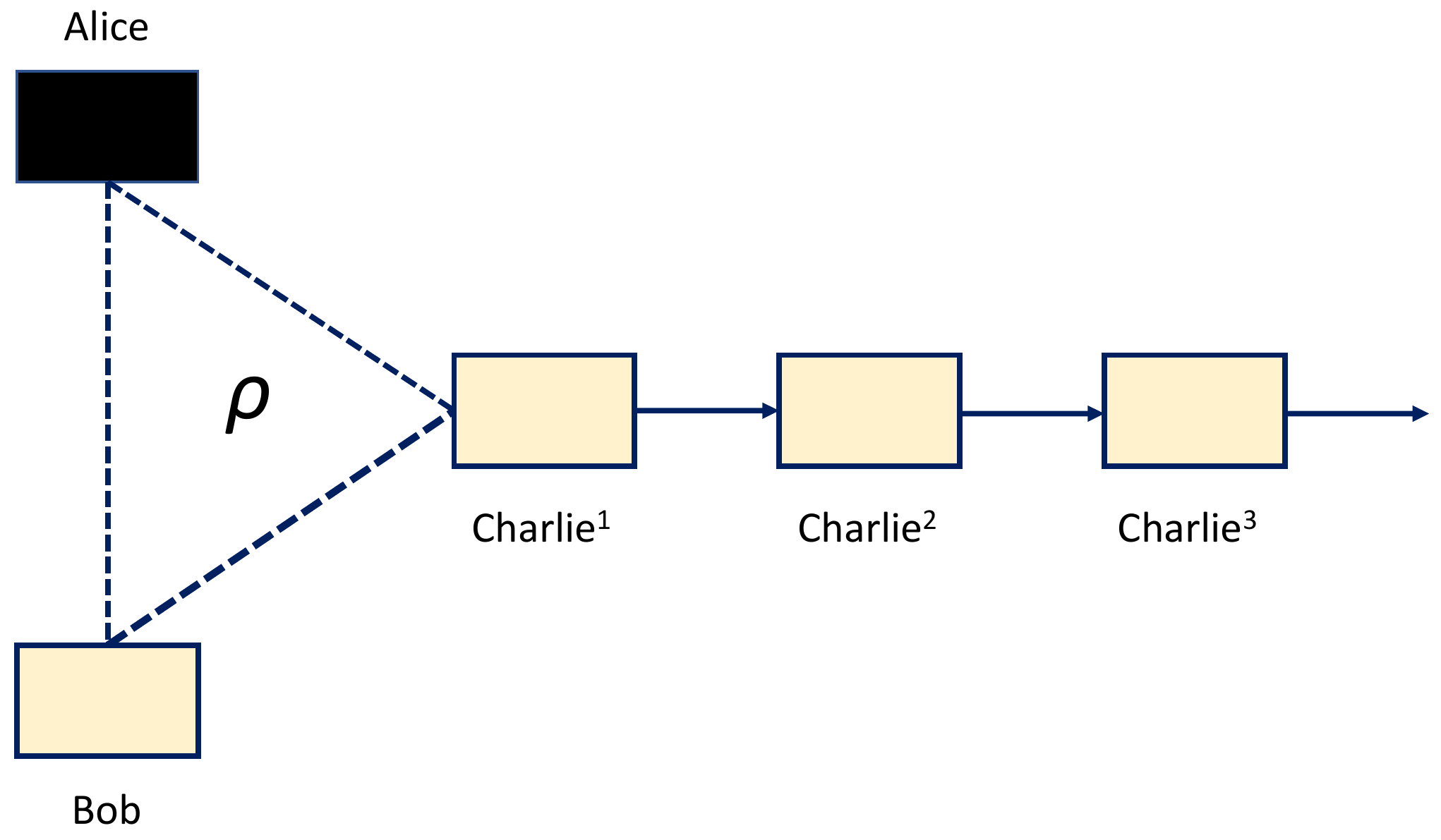}
\caption{(Color Online) Sequential detection of genuine tripartite EPR steering with multiple Charlies in the scenario where Alice tries to demonstrate genuine tripartite    
($1 \rightarrow 2$) steering to Bob-Charlie. Three spatially separated spin-$\frac{1}{2}$ particles,  prepared in the three-qubit state $\rho$,  are shared between a single Alice, a single Bob and multiple Charlies (Charlie$^1$, Charlie$^2$, $\cdots$, Charlie$^n$). }
\label{fig3}
\end{figure}

\begin{figure}[t!]
\centering
\includegraphics[scale=0.4]{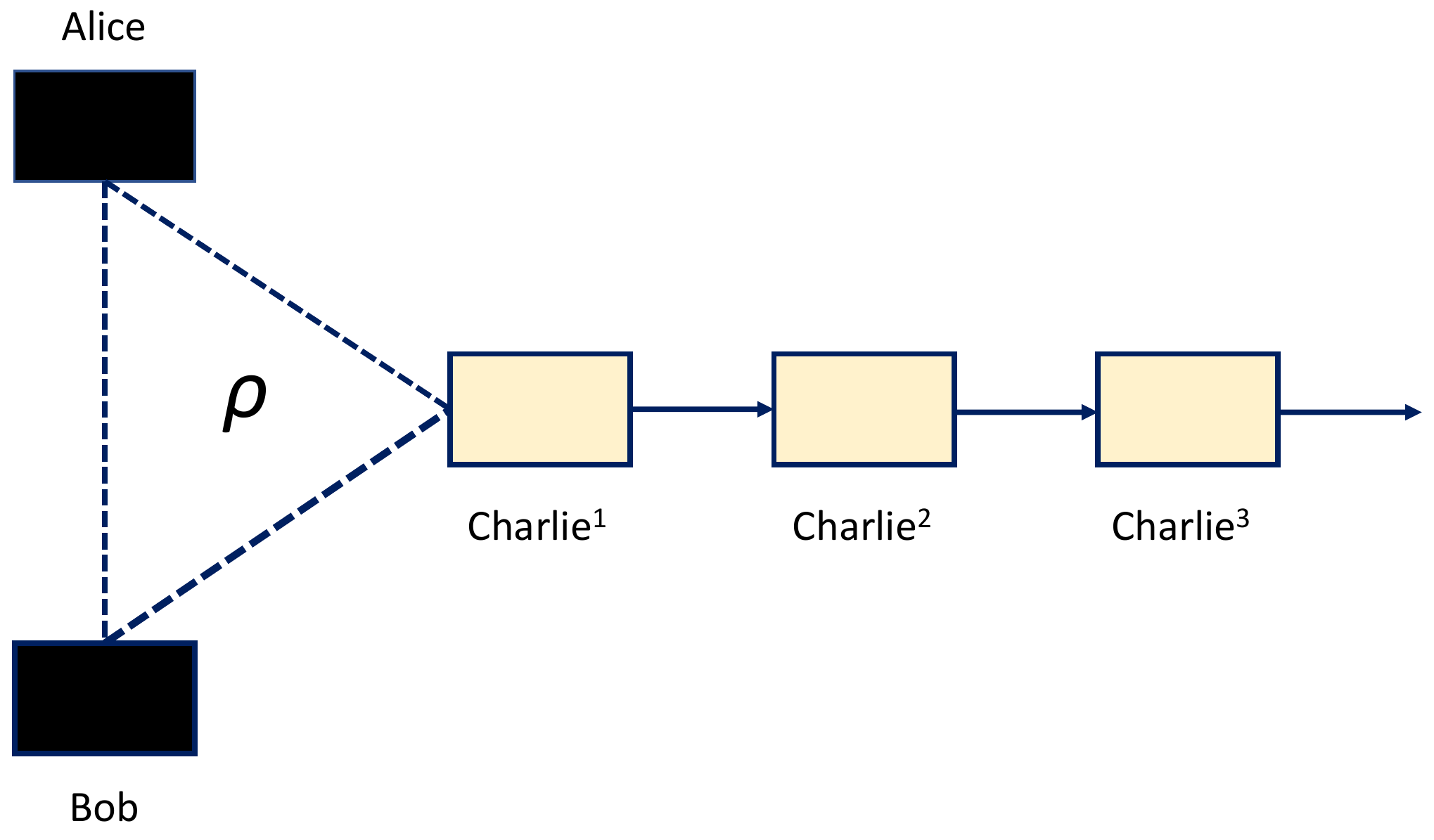}
\caption{(Color Online) Sequential detection of genuine tripartite EPR steering with multiple  Charlies in the scenario where both Alice and Bob try to demonstrate genuine tripartite ($2 \rightarrow 1$) steering to Charlie. Three spatially separated spin-$\frac{1}{2}$ particles,  prepared in the three-qubit state $\rho$,  are shared between a single Alice, a single Bob and multiple Charlies (Charlie$^1$, Charlie$^2$, $\cdots$, Charlie$^n$).  }
\label{fig4}
\end{figure}

The aforementioned scenarios are depicted in Figures \ref{fig1}, \ref{fig2}, \ref{fig3}, and \ref{fig4}, respectively. 

Next, let us briefly discuss the unsharp measurement formalism used in this paper (For details, see \cite{sygp,majumdar,sas}). In a sharp projective measurement, one obtains the maximum amount of information at the cost of maximum disturbance to the state. On the other hand, in our scenario, Alice$^m$ (or, Charlie$^m$) passes on the respective particle to Alice$^{m+1}$ (or, Charlie$^{m+1}$) after performing suitable measurement. Hence, in this case, Alice$^m$ (or, Charlie$^m$) needs to demonstrate genuine tripartite EPR steering by disturbing the state minimally so that Alice$^{m+1}$ (or, Charlie$^{m+1}$) can again demonstrate genuine tripartite EPR steering. This can be achieved by unsharp measurement   \cite{sygp} which may be characterized by two real parameters: the quality factor $F$ and the precision $G$ of the measurement. The quality factor $F$ quantifies the extend to which the initial state of the system (to be measured) remains undisturbed during  the measurement process and the precision $G$ quantifies the information gain due to the measurement. In case of projective measurement, $F =0$, $G=1$. For dichotomic measurements on a qubit system, the optimal trade-off relation between precision $G$ and quality factor $F$ is given by, $F^2 + G^2 =1$ \cite{sygp}. In other words, for dichotomic measurements on a qubit system, satisfying the condition: $F^2 +G^2 = 1$ implies that the disturbance is minimized for any particular information gain.

The above optimal trade-off relation between information gain and quality factor is achieved under unsharp measurement formalism \cite{majumdar,sas}. Unsharp measurement \cite{pb1,pb2} is one particular class of  positive operator valued measurements (POVM) \cite{pb1,pb2}. A POVM is nothing but a set of positive operators that add to identity, i. e.,  $E \equiv \{ E_i | \sum_i E_i = \mathbb{I}, 0 <E_i \leq \mathbb{I} \, \forall \, i \}$. Here, each of the effect operators $E_i$ determines the probability  $\text{Tr}[\rho E_{i}]$ of obtaining the $i^{\text{th}}$ outcome (here $\rho$ is the state of the system on which the measurement is performed). 
 For example, consider a dichotomic observable $O$ $=$ $\Pi_{+} - \Pi_{-}$ with outcomes $+1$ and $-1$, where $\Pi_{+}$ and $\Pi_{-}$ denote the projectors associated with the outcomes $+1$ and $-1$ respectively, with $\Pi_{+} +  \Pi_{-} = \mathbb{I}$ and $\Pi_{\pm}^2 = \Pi_{\pm}$. Given the observable $O$, a dichotomic unsharp observable  $O^{\lambda} = E^{\lambda}_+ - E^{\lambda}_-$ \cite{umn1,umn2} can be defined which is associated with the sharpness parameter $\lambda \in (0, 1]$. Here, the effect operators $E^{\lambda}_{\pm}$ are given by, 
\begin{equation}
E^\lambda_{\pm} = \lambda \, \Pi_{\pm} + (1-\lambda) \frac{\mathbb{I}_2}{2}.
\end{equation} 
This is obtained by mixing projective measurements with white noise. The probability of getting the outcomes $+1$ and $-1$, when the above unsharp measurement is performed on the state $\rho$, are given by $\text{Tr}[\rho E^\lambda_{+}]$ and $\text{Tr}[\rho E^\lambda_{-}]$ respectively. Using the generalized von Neumann-L\"{u}ders transformation rule \cite{pb1}, the states after the measurements, when the outcomes $+1$ and $-1$ occurs, are given by, $\dfrac{\sqrt{E^\lambda_{+}} \rho \sqrt{E^\lambda_{+}}}{\text{Tr}[E^\lambda_{+} \rho]}$ and $\dfrac{\sqrt{E^\lambda_{-}} \rho \sqrt{E^\lambda_{-}}}{\text{Tr}[E^\lambda_{-} \rho]}$ respectively. 

For the von Neumann-L\"{u}ders transformation rule in the  unsharp measurement formalism, it was shown \cite{majumdar} that the quality factor $F$ and the precision $G$ are given by, $F = \sqrt{1-\lambda^2}$ and $G = \lambda$. Hence, the optimal trade-off relation between information gain and disturbance, $F^2 + G^2 =1$, for qubits is compatible with the above unsharp measurement formalism \cite{majumdar,sas}. In other words, the unsharp measurement formalism along with the von Neumann-L\"{u}ders transformation rule provides the largest amount of information for a given amount of disturbance created on the state due to the measurement. 

In our study, we  consider that multiple Alices (in Scenario A) or multiple Charlies (in Scenario B) in the sequence, except for the last one, perform unsharp measurements.  We obtain the violations of the inequalities (\ref{GHZ1}, \ref{GHZ2}, \ref{W1}, \ref{W2}) by calculating the expectation values associated with different unsharp observables. The expectation value of $O^{\lambda}$ for a given $\rho$ can be defined in the following way \cite{umn1,umn2},
\begin{align}
\langle O^{\lambda} \rangle &= \text{Tr}[\rho E^\lambda_{+}] - \text{Tr}[\rho E^\lambda_{-}] \nonumber \\
&= \text{Tr}[\rho (E^\lambda_{+} -  E^\lambda_{-})] \nonumber \\
&= \lambda \langle O \rangle.
\end{align}
Here, $\langle O \rangle$ is the expectation value of the observable $O$ under projective measurements. Hence, from the probabilities (i.e., $\text{Tr}[\rho E^\lambda_{\pm}]$) of obtaining the outcomes $\pm 1$ under unsharp measurements, one can evaluate the expectation value of $O^{\lambda}$. 

Experimental implementation of such unsharp measurement formalism has  been demonstrated using trapped-ion systems \cite{une1}. Recently, the unsharp measurement formalism has also been implemented in photonic systems \cite{exp1,exp2,Choi2020,exp4,rac1} for the purpose of sequential measurement scenario used in the present context. Using an appropriate interferometer one can realize  unsharp measurements, where the sharpness parameter can be controlled by fine-tuning the arrangement of various components in the interferometer.

\section{Steering by multiple observers}\label{s3}

Using the formalism discussed in the earlier section we are now in a position to explore the maximum number of parties that can demonstrate  genuine tripartite EPR steering in the previously mentioned Scenarios A and B. 

\subsection{Multiple Alices performing sequential measurements}\label{mm}
 
 In this subsection, we  address two specific questions dealing with the sequential detection of genuine tripartite EPR steering in Scenario A mentioned earlier, i.e., when multiple Alices perform sequential measurements on the first particle, single Bob and single Charlie  perform measurements on the second and third particle respectively. Let Bob perform dichotomic projective measurement of the spin component observable  in the direction $\hat{y}_0$, or  $\hat{y}_1$, or $\hat{y}_2$. Charlie performs dichotomic projective measurement of the spin component observable  in the direction $\hat{z}_0$, or $\hat{z}_1$, or $\hat{z}_2$.  Alice$^m$ (where $m$ $\in \{1, 2, \cdots, n\}$) performs dichotomic unsharp measurement of the spin component observable  in the direction $\hat{x}_0^m$, or $\hat{x}_1^m$, or $\hat{x}_2^m$. The outcomes of each measurement is $\pm1$.

The projectors associated with Bob's sharp spin component measurement in the direction $\hat{y}_j$ (with $j$ $\in$ $\{0,1,2\}$) are given by, $\Pi_{b|\hat{y}_j} = \dfrac{\mathbb{I}_2+b \, \hat{y}_j \cdot \vec{\sigma}}{2}$ (with $b$ $\in$ $\{+1, -1\}$ being the outcome of Bob's sharp measurement). Similarly, the projectors associated with Charlie's sharp spin component measurement in the direction $\hat{z}_k$ (with $k$ $\in$ $\{0,1,2\}$) can be written as $\Pi_{c|\hat{z}_k} = \dfrac{\mathbb{I}_2+c \, \hat{z}_k \cdot \vec{\sigma}}{2}$ (with $c$ $\in$ $\{+1, -1\}$ being the outcome of Charlie's sharp measurement). Here $\vec{\sigma}$ = $(\sigma_1, \sigma_2, \sigma_3)$ is a vector composed of three
Pauli matrices. The directions $\hat{y}_j$ and $\hat{z}_k$ can be expressed as,
\begin{equation}
\label{bobdir}
\hat{y}_j = \sin \theta^{y}_j \cos \phi^{y}_j \hat{X} + \sin \theta^{y}_j \sin \phi^{y}_j \hat{Y} + \cos \theta^{y}_j \hat{Z},
\end{equation}
and 
\begin{equation}
\label{charliedir}
\hat{z}_k = \sin \theta^{z}_k \cos \phi^{z}_k \hat{X} + \sin \theta^{z}_k \sin \phi^{z}_k \hat{Y} + \cos \theta^{z}_k \hat{Z},
\end{equation}
where $j, k \in \{0, 1, 2\}$; $0 \leq \theta^{y}_j  \leq \pi$; $0 \leq \phi^{y}_j  \leq 2 \pi$; $0 \leq \theta^{z}_k  \leq \pi$; $0 \leq \phi^{z}_k  \leq 2 \pi$. $\hat{X}$, $\hat{Y}$, $\hat{Z}$ are three orthogonal unit vectors in Cartesian coordinates. 

The effect operators associated with Alice$^m$'s ($m$ $\in$ $\{1, 2, \cdots, n \}$) unsharp measurement of spin component observable in the direction $\hat{x}^m_i$ (with $i$ $\in$ $\{0,1,2\}$) are given by,
\begin{equation}
E^{\lambda_m}_{a^m|\hat{x}^m_i} = \lambda_{m}\frac{\mathbb{I}_2+a^{m} \hat{x}^m_i \cdot \vec{\sigma}}{2}+(1-\lambda_{m})\frac{\mathbb{I}_2}{2},
\end{equation}
with $a^m$ $\in$ $\{+1, -1\}$ being the outcome of Alice$^m$'s unsharp measurement and $\lambda_m$ ($0 < \lambda_m \leq 1$) is the sharpness parameter corresponding to Alice$^m$'s unsharp measurement. For a sequence of $n$ Alices, the measurements of Alice$^n$ will be sharp, i.e., $\lambda_n$ = $1$. The direction $\hat{x}^m_i$ is given by,
\begin{equation}
\label{charliemdir}
\hat{x}^m_i = \sin \theta^{x^m}_i \cos \phi^{x^m}_i \hat{X} + \sin \theta^{x^m}_i \sin \phi^{x^m}_i \hat{Y} + \cos \theta^{x^m}_i \hat{Z},
\end{equation}
where $i \in \{0, 1, 2\}$; $0 \leq \theta^{x^m}_i  \leq \pi$; $0 \leq \phi^{x^m}_i  \leq 2 \pi$.

There are various types of correlations appearing in the inequalities (\ref{GHZ1}, \ref{GHZ2}, \ref{W1}, \ref{W2}). In the following, we compute these correlations between Alice$^m$, Bob and Charlie. 

The joint probability distribution of occurrence of the outcomes $a^1$, $b$, $c$, when Alice$^1$ performs unsharp measurement of spin component observable along the direction $\hat{x}^1_i$, and Bob and Charlie perform projective measurements of spin component observables along the directions $\hat{y}_j$ and $\hat{z}_k$ respectively on the shared tripartite state $\rho$, is given by,
\begin{align}
&P(a^1, b, c|\hat{x}^1_i, \hat{y}_j, \hat{z}_k) \nonumber \\
&=\text{Tr}\Bigg[\Bigg\{ E^{\lambda_1}_{a^1|\hat{x}^1_i} \otimes \frac{\mathbb{I}_2 + b \hat{y}_j \cdot \vec{\sigma}}{2} \otimes \frac{\mathbb{I}_2+ c \hat{z}_k \cdot \vec{\sigma}}{2}  \Bigg\}  \cdot \rho \Bigg].
\end{align}
In this case, the correlation function between Alice$^1$, Bob and Charlie can be written as 
\begin{equation}
\langle x^1_i \, y_j \, z_k \rangle =\sum_{a^1 = -1}^{ +1} \sum_{b =  -1}^{ +1} \sum_{c = -1}^{ +1} a^1 \, b \, c \, P(a^1, b, c|\hat{x}^1_i, \hat{y}_j, \hat{z}_k).
\end{equation} 

After performing unsharp measurement, Alice$^1$ passes her particle to Alice$^2$. The unnormalized post-measurement reduced state at Alice$^2$'s end, after Alice$^1$ gets the outcome $a^1$ by performing unsharp measurement of spin component observable along the direction $\hat{x}^1_i$ and Bob  and Charlie get the outcomes $b$ and $c$ by performing sharp measurements of spin component observables along the directions $\hat{y}_j$ and $\hat{z}_k$ respectively, is given by,
\begin{align}
\rho_{un}^{A^2} =& \text{Tr}_{B C} \Bigg[ \Bigg\{ \sqrt{E^{\lambda_1}_{a^1|\hat{x}^1_i}}  \otimes \frac{\mathbb{I}_2 + b \hat{y}_j \cdot \vec{\sigma}}{2} \otimes \frac{\mathbb{I}_2+ c \hat{z}_k \cdot \vec{\sigma}}{2} \Big\}  \nonumber \\ 
& \cdot \rho \cdot \Bigg\{ \sqrt{E^{\lambda_1}_{a^1|\hat{x}^1_i}}  \otimes \frac{\mathbb{I}_2 + b \hat{y}_j \cdot \vec{\sigma}}{2} \otimes \frac{\mathbb{I}_2+ c \hat{z}_k \cdot \vec{\sigma}}{2} \Big\} \Bigg],
\end{align}
where,
\begin{align}
\sqrt{E^{\lambda_1}_{a^1|\hat{x}^1_i}} &= \sqrt{\dfrac{1+\lambda_1}{2}} \Bigg( \dfrac{\mathbb{I}_2 + a^1 \hat{x}^1_i \cdot \vec{\sigma}}{2} \Bigg) \nonumber \\
& +\sqrt{\dfrac{1- \lambda_1}{2}} \Bigg( \dfrac{\mathbb{I}_2 - a^1 \hat{x}^1_i \cdot \vec{\sigma}}{2} \Bigg).
\end{align}
In order to get the reduced state, the partial trace has been taken over the subsystems of Bob and Charlie.

Now Alice$^2$ again performs unsharp measurement (associated with sharpness parameter $\lambda_{2}$) of spin component observable along the direction $\hat{x}^2_l$ on the reduced state $\rho_{un}^{A^2}$ and gets the outcome $a^2$. The joint probability distribution of occurrence of the outcomes $a^1$, $a^2$, $b$, $c$, when Alice$^1$, Alice$^2$ perform unsharp measurements of spin component observables along the directions $\hat{x}^1_i$, $\hat{x}^2_l$ respectively and Bob, Charlie perform projective measurements of spin component observables along the directions $\hat{y}_j$ and $\hat{z}_k$ respectively, is given by,
\begin{equation}
P(a^1, a^2, b, c|\hat{x}^1_i, \hat{x}^2_l, \hat{y}_j, \hat{z}_k) = \text{Tr} \Big[ E^{\lambda_2}_{a^2|\hat{x}^2_l} \cdot \rho^{A^2}_{un} \Big].
\end{equation}
 From this expression, the joint probability of obtaining the outcomes $a^2$, $b$, $c$ by Alice$^2$, Bob, Charlie, respectively, can be calculated as,
\begin{align}
& P(a^2, b, c|\hat{x}^1_i, \hat{x}^2_l, \hat{y}_j, \hat{z}_k) \nonumber \\
&= \sum_{a^1 = -1}^{ +1} P(a^1, a^2, b, c|\hat{x}^1_i, \hat{x}^2_l, \hat{y}_j, \hat{z}_k).
\end{align}


Let $\langle x^2_l \, y_j \, z_k \rangle_{x^1_i}$ denote the correlation between Alice$^2$, Bob and Charlie, when Alice$^1$, Alice$^2$ perform unsharp measurements of spin component observables along the directions $\hat{x}^1_i$, $\hat{x}^2_l$ respectively and Bob, Charlie perform projective measurements of spin component observables along the directions $\hat{y}_j$ and $\hat{z}_k$ respectively. The expression for $\langle x^2_l \, y_j \, z_k \rangle_{x^1_i} $ can be obtained as,
\begin{align}
& \langle x^2_l \, y_j \, z_k \rangle_{x^1_i} \nonumber \\
&= \sum_{a^2 = -1}^{ +1} \sum_{b =  -1}^{ +1} \sum_{c = -1}^{ +1} a^2 \, b \, c \, P(a^2, b, c|\hat{x}^1_i, \hat{x}^2_l, \hat{y}_j, \hat{z}_k).
\end{align}
Since Alice$^2$ is ignorant about the choice of the measurement setting of Alice$^1$, the above correlation has to be averaged over the three possible measurement settings of Alice$^1$ (unsharp measurement of spin component observables in the directions $\{ \hat{x}^1_0, \hat{x}^1_1, \hat{x}^1_2 \}$). This average correlation function between Alice$^2$, Bob and Charlie is given by,
\begin{equation}
\langle x^2_l \, y_j \, z_k \rangle_{\text{av}} = \sum_{i = 0,1,2} \langle x^2_l \, y_j \, z_k \rangle_{x^1_i} \, P(\hat{x}^1_i).
\label{avcor}
\end{equation}
Here $P(\hat{x}^1_i)$ is the probability of Alice$^1$'s unsharp measurement of spin component observable in the direction $\hat{x}^1_i$ ($i \in \{ 0, 1, 2 \}$). Since, we restrict ourselves to unbiased input scenario, all the three measurement settings for Alice$^1$ are equally probable, i.e.,  $P(\hat{x}^1_0)$ = $P(\hat{x}^1_1)$ = $P(\hat{x}^1_2)$ = $\frac{1}{3}$. 

Using this general expression (\ref{avcor}) for the average three-party correlation function, the terms appearing in inequalities (\ref{GHZ1}, \ref{GHZ2}, \ref{W1}, \ref{W2}) can be easily calculated for detection of genuine tripartite EPR steering by Alice$^2$. 
Note that there are several two-party correlation functions and one-party expectation values on the left hand sides of inequalities (\ref{GHZ1}, \ref{GHZ2}, \ref{W1}, \ref{W2}). These can be calculated using the above-mentioned approach invoking the no-signalling condition between the observers at three different wings. For example, the average correlation function between Alice$^2$ and Bob can be calculated as follows.

The joint probability distribution of occurrence of the outcomes $a^2$, $b$, when Alice$^1$, Alice$^2$ perform unsharp measurements of spin component observables along the directions $\hat{x}^1_i$, $\hat{x}^2_l$ respectively and Bob performs projective measurement of spin component observable along the direction $\hat{y}_j$, is given by,
\begin{equation}
P(a^2, b|\hat{x}^1_i, \hat{x}^2_l, \hat{y}_j) = \sum_{c=-1}^{+1} P(a^2, b, c|\hat{x}^1_i, \hat{x}^2_l, \hat{y}_j, \hat{z}_k).
\end{equation}
Here, we have used the no-signalling condition between Alice$^2$, Bob and Charlie. 
In the above case, the average two-party correlation function $\langle x^2_l \, y_j  \rangle_{\text{av}}$ between Alice$^2$ and Bob is given by,
\begin{align}
& \langle x^2_l \, y_j  \rangle_{\text{av}} \nonumber \\
&= \sum_{i = 0,1,2} \Bigg[ \sum_{a^2 = -1}^{ +1} \sum_{b =  -1}^{ +1} a^2 \, b  \, P(a^2, b|\hat{x}^1_i, \hat{x}^2_l, \hat{y}_j) \Bigg] \, P(\hat{x}^1_i).
\label{avcor2}
\end{align}
Following the above-mentioned approach, each term appearing on the left hand sides of the inequalities (\ref{GHZ1}, \ref{GHZ2}, \ref{W1}, \ref{W2}) in the context of  Alice$^m$, Bob and Charlie can be calculated. In the following, we consider the initial state is a three-qubit GHZ state.

\subsubsection{When the three-qubit GHZ state is initially shared} \label{sub1}
Let us consider that the three-qubit GHZ-state \cite{GHZ} given by $\rho_{\text{GHZ}} = | \psi_{\text{GHZ}} \rangle \langle \psi_{\text{GHZ}} |$  is shared between multiple Alices, Bob and Charlie, where
\begin{equation}
|\psi_{\text{GHZ}} \rangle = \frac{1}{\sqrt{2}} ( |000 \rangle + | 111 \rangle )
\label{ghz}
\end{equation}
Here $|0\rangle$ and $|1\rangle$ denotes two mutually orthonormal states in $\mathbb{C}^2$. Here, multiple Alices perform sequential unsharp measurements.

 At first, we will find out the maximum number of Alices, who can steer Bob-Charlie. This will be probed through the quantum violation of the inequality (\ref{GHZ1}).


 We start by finding out  whether Alice$^1$ and Alice$^2$ can sequentially demonstrate genuine tripartite steering in this case. In other words, we will find out whether Alice$^1$ and Alice$^2$ can sequentially violate the inequality (\ref{GHZ1}) with single Bob and single Charlie. In this case, the measurements of the final Alice in the sequence,  i.e., Alice$^2$ will be sharp ($\lambda_2 = 1$), and the measurements of Alice$^1$ will be unsharp. We observe that, for example, when  Alice$^1$ gets $G_1 = -0.10$, then Alice$^2$ gets $G_1 = -0.55$. This happens for the following choices of measurement settings by Alice$^1$ and Alice$^2$:  $( \theta^{x^1}_0$, $\phi^{x^1}_0$, $\theta^{x^1}_1$, $\phi^{x^1}_1$, $\theta^{x^1}_2$, $\phi^{x^1}_2$,  $\theta^{x^2}_0$, $\phi^{x^2}_0$, $\theta^{x^2}_1$, $\phi^{x^2}_1$, $\theta^{x^2}_2$, $\phi^{x^2}_2$ $)$ $\equiv$ $( \frac{\pi}{2}$, $0$, $\frac{\pi}{2}$, $\frac{\pi}{2}$, $0$, $0$,  $\frac{\pi}{2}$, $0$, $\frac{\pi}{2}$, $\frac{\pi}{2}$, $0
 $, $0$ $)$, with $\lambda_1 = 0.627$. Note that, here the choices of measurement settings by Bob and Charlie are not mentioned as they are already specified in the inequality (\ref{GHZ1}). Hence, we can conclude that Alice$^1$ and Alice$^2$ can genuinely steer Bob and Charlie when the GHZ state is initially shared.

Next, we ask whether Alice$^1$, Alice$^2$ and Alice$^3$ can sequentially violate the inequality (\ref{GHZ1}) with single Bob and single Charlie. In this case, the measurements of the final Alice, i.e., Alice$^3$ are sharp ($\lambda_3 = 1$), and the measurements of Alice$^1$ and Alice$^2$ are unsharp. We observe that, when Alice$^1$ gets $G_1 = -0.10$ and Alice$^2$ gets $G_1 = -0.10$, then Alice$^3$ gets $G_1 = -0.18$.  This happens for the following choices of measurement settings by Alice$^1$, Alice$^2$ and Alice$^3$:  $( \theta^{x^1}_0$, $\phi^{x^1}_0$, $\theta^{x^1}_1$, $\phi^{x^1}_1$, $\theta^{x^1}_2$, $\phi^{x^1}_2$,  $\theta^{x^2}_0$, $\phi^{x^2}_0$, $\theta^{x^2}_1$, $\phi^{x^2}_1$, $\theta^{x^2}_2$, $\phi^{x^2}_2$, $\theta^{x^3}_0$, $\phi^{x^3}_0$, $\theta^{x^3}_1$, $\phi^{x^3}_1$, $\theta^{x^3}_2$, $\phi^{x^3}_2$ $)$ $\equiv$ $( \frac{\pi}{2}$, $0$, $\frac{\pi}{2}$, $\frac{\pi}{2}$, $0$, $0$,  $\frac{\pi}{2}$, $0$, $\frac{\pi}{2}$, $\frac{\pi}{2}$, $0$, $0$,  $\frac{\pi}{2}$, $0$, $\frac{\pi}{2}$, $\frac{\pi}{2}$, $0$, $0$ $)$, with $\lambda_1 = 0.627$ and $\lambda_2 = 0.736$. Hence, Alice$^1$, Alice$^2$ and Alice$^3$ can genuinely steer Bob and Charlie when the GHZ state is initially shared.

Now we  investigate whether Alice$^1$, Alice$^2$, Alice$^3$ and Alice$^4$ can sequentially violate the inequality (\ref{GHZ1}) with single Bob and single Charlie. Here, the measurements of Alice$^4$ are sharp ($\lambda_4 = 1$), and the measurements of all other Alices are unsharp. In this case, we observe that for any choices of measurement settings by multiple Alices and for any permissible values of $\lambda_1$, $\lambda_2$, $\lambda_3$, Alice$^1$, Alice$^2$, Alice$^3$ and Alice$^4$ cannot sequentially violate the inequality (\ref{GHZ1}). These results are summarized in Table \ref{tab1}.    The permissible range of each $\lambda_m$ depends on the values $\lambda_1$, $\lambda_2$, $\cdots$, $\lambda_{m-1}$. In the table, we have presented the permissible range of each $\lambda_m$ for the minimum permissible value of each $\lambda_1$, $\lambda_2$, $\cdots$, $\lambda_{m-1}$. The permissible range of $\lambda_m$ will be smaller than this if we take other value $\lambda_i$ > $\lambda_i^{\text{min}}$ $\forall$ $i < m$, and the maximum number of Alices may get reduced. It is to be noted here that Alice$^4$ may obtain quantum mechanical violation of the  inequality (\ref{GHZ1}) if the sharpness parameter of any previous Alice is too small not to get a violation. In fact, any three Alices (at most) can sequentially demonstrate genuine tripartite steering to Bob-Charlie by  violating the inequality (\ref{GHZ1}).

{\centering
	\begin{table}[t]\footnotesize
			\begin{tabular}{|c | c || c |} 
				\hline 
				& \textbf{Genuine tripartite} & \textbf{Genuine tripartite} \\
				& \textbf{($1 \rightarrow 2$) EPR steering} & \textbf{($2 \rightarrow 1$) EPR steering} \\
				& \textbf{from Alice$^m$} & \textbf{from Alice$^m$-Bob} \\
				& \textbf{to Bob-Charlie} & \textbf{to Charlie} \\
				\hline
				{\bf Alice$^m$} & {\bf Permissible $\lambda_{m}$ range}  & {\bf Permissible $\lambda_{m}$ range}  \\ [0.5ex] 
				\hline
				\hline
				Alice$^1$ &  1 $\geq \lambda_1 > \lambda_1^{\text{min}}$ = 0.577 &  1 $\geq \lambda_1 > \lambda_1^{\text{min}}$ = 0.584  \\ 
				\hline
				Alice$^2$ &1 $\geq \lambda_2 > \lambda_2^{\text{min}}$ = 0.658  & 1 $\geq \lambda_2 > \lambda_2^{\text{min}}$ = 0.668 \\
				& when $\lambda_i = \lambda_i^{\text{min}}$ $\forall$ $i < 2$ & when $\lambda_i = \lambda_i^{\text{min}}$ $\forall$ $i < 2$ \\
				\hline
				Alice$^3$ &1 $\geq \lambda_3 > \lambda_3^{\text{min}}$ = 0.787  & 1 $\geq \lambda_3 > \lambda_3^{\text{min}}$ = 0.805  \\
				& when $\lambda_i = \lambda_i^{\text{min}}$ $\forall$ $i < 3$ & when $\lambda_i = \lambda_i^{\text{min}}$ $\forall$ $i < 3$ \\
				\hline
				Alice$^4$ & No valid range for $\lambda_4$   & No valid range for $\lambda_4$\\
				&  for any $\lambda_i$ with $i<4$ &  for any $\lambda_i$ with $i<4$\\ [1ex]
				\hline
			\end{tabular}
			\caption{The permissible ranges of the sharpness parameters $\lambda_m$ (where $0 < \lambda_m \leq 1$) of Alice$^m$ in order to demonstrate genuine steering when the  three-qubit GHZ state is initially shared.}
			\label{tab1}
	\end{table}
}

 Next, we obtain the maximum number of Alices, who, along with the single Bob, can genuinely steer Charlie's particle. This will be probed through the quantum violation of the inequality (\ref{GHZ2}). Following a similar approach,  in this case, we find that the maximum number of Alices is three. This result is also summarized in Table \ref{tab1}. 

Thus, for the GHZ state, we get the maximum numbers of Alices to be three  in both the cases (i.e., in Alice$^m$ to Bob-Charlie genuine tripartite ($1 \rightarrow 2$) steering case, or in Alice$^m$-Bob to Charlie genuine tripartite ($2 \rightarrow 1$) steering case). However, the allowed range of the sharpness parameters is larger in the  ($1 \rightarrow 2$) steering cases.

\subsubsection{When the three-qubit W state is initially shared}
Here, let us consider that the three-qubit W state $\rho_{\text{W}} = | \psi_{\text{W}} \rangle \langle \psi_{\text{W}} |$ is initially shared between multiple Alices, single Bob and single Charlie, where
\begin{equation}
| \psi_{\text{W}} \rangle = \frac{1}{\sqrt{3}}(| 001\rangle + |010\rangle + |100 \rangle)
\label{wstate}
\end{equation} 

Let us now explore the maximum number of Alices who can  sequentially demonstrate genuine tripartite steering to Bob-Charlie. Here we  use the genuine tripartite EPR steering inequality (\ref{W1}). Following the approach mentioned in Section \ref{sub1}, we observe that at most two Alices can sequentially demonstrate genuine tripartite steering to Bob-Charlie by  violating the inequality (\ref{W1}).

 Next, let us find out the maximum number of Alices, who, along with the single Bob, can  sequentially demonstrate genuine tripartite steering to Charlie through the quantum  violation of the genuine tripartite EPR steering inequality (\ref{W2}). In this case also, the maximum number of Alices turns out to be two.

These results are summarized in Table \ref{tab2}. For the W state too, the maximum number of Alices are the same  in both the cases- from Alice$^m$ to Bob-Charlie genuine tripartite ($1 \rightarrow 2$) steering, and from Alice$^m$-Bob to Charlie genuine tripartite ($2 \rightarrow 1$) steering case. Again, the allowed range of the sharpness parameters is larger in the  ($1 \rightarrow 2$) steering cases.

{\centering
	\begin{table}[t]\footnotesize
			\begin{tabular}{|c | c || c |} 
				\hline 
				& \textbf{Genuine tripartite} & \textbf{Genuine tripartite} \\
				& \textbf{($1 \rightarrow 2$) EPR steering} & \textbf{($2 \rightarrow 1$) EPR steering} \\
				& \textbf{from Alice$^m$} & \textbf{from Alice$^m$-Bob} \\
				& \textbf{to Bob-Charlie} & \textbf{to Charlie} \\
				\hline
				{\bf Alice$^m$} & {\bf Permissible $\lambda_{m}$ range}  & {\bf Permissible $\lambda_{m}$ range}  \\ [0.5ex] 
				\hline
				\hline
				Alice$^1$ & 1 $\geq \lambda_1 > \lambda_1^{\text{min}}$ = 0.588  & 1 $\geq \lambda_1 > \lambda_1^{\text{min}}$ =  0.678  \\ 
				\hline
				Alice$^2$ & 1 $\geq \lambda_2 > \lambda_2^{\text{min}}$ = 0.674  & 1 $\geq \lambda_2 > \lambda_2^{\text{min}}$ = 0.823 \\
				& when $\lambda_i = \lambda_i^{\text{min}}$ $\forall$ $i < 2$ & when $\lambda_i = \lambda_i^{\text{min}}$ $\forall$ $i < 2$ \\
				\hline
				Alice$^3$ & No valid range for $\lambda_3$ & No valid range for $\lambda_3$  \\
				& for any $\lambda_i$ with $i<3$ & for any $\lambda_i$ with $i<3$ \\[1ex]
				\hline
			\end{tabular}
			\caption{The permissible ranges of the sharpness parameters $\lambda_m$ (where $0 < \lambda_m \leq 1$) of Alice$^m$ in order to demonstrate genuine steering when the three-qubit W state  is initially shared.}
			\label{tab2}
	\end{table}
}


\subsection{Multiple Charlies performing sequential measurements}
Now, we address the specific two questions mentioned for Scenario B. Here, a tripartite three-qubit state $\rho$ (either GHZ state or W state) consisting of three spin-$\frac{1}{2}$ particles is initially shared among Alice, Bob and multiple Charlies. Alice performs dichotomic sharp measurement of spin component observable on first particle in the direction $\hat{x}_0$, or  $\hat{x}_1$, or $\hat{x}_2$. Bob performs dichotomic sharp measurement of spin component observable on second particle in the direction $\hat{y}_0$, or $\hat{y}_1$, or $\hat{y}_2$.  Charlie$^m$ (where $m$ $\in \{1, 2, \cdots, n\}$) performs dichotomic unsharp measurement (associated with sharpness parameter $\lambda_m$) of spin component observable on third particle in the direction $\hat{z}_0^m$, or $\hat{z}_1^m$, or $\hat{z}_2^m$. The outcomes of each measurement are $\pm1$. Here
\begin{equation}
\label{alice2dir}
\hat{x}_i = \sin \theta^{x}_i \cos \phi^{x}_i \hat{X} + \sin \theta^{x}_i \sin \phi^{x}_i \hat{Y} + \cos \theta^{x}_i \hat{Z},
\end{equation}
\begin{equation}
\label{bob2dir}
\hat{y}_j = \sin \theta^{y}_j \cos \phi^{y}_j \hat{X} + \sin \theta^{y}_j \sin \phi^{y}_j \hat{Y} + \cos \theta^{y}_j \hat{Z},
\end{equation}
and 
\begin{equation}
\label{charlie2mdir}
\hat{z}^m_k = \sin \theta^{z^m}_k \cos \phi^{z^m}_k \hat{X} + \sin \theta^{z^m}_k \sin \phi^{z^m}_k \hat{Y} + \cos \theta^{z^m}_k \hat{Z},
\end{equation}
where $i, j, k \in \{0, 1, 2\}$; $0 \leq \theta^{x}_i  \leq \pi$; $0 \leq \phi^{x}_i  \leq 2 \pi$; $0 \leq \theta^{y}_j  \leq \pi$; $0 \leq \phi^{y}_j  \leq 2 \pi$; $0 \leq \theta^{z^m}_k  \leq \pi$; $0 \leq \phi^{z^m}_k  \leq 2 \pi$.

As in earlier cases, genuine tripartite steering is probed through the quantum violations of the inequalities (\ref{GHZ1}, \ref{GHZ2}, \ref{W1}, \ref{W2}). Here, the correlation functions and expectation values can be calculated using the technique described in Section \ref{mm}, with only the role of Alice and Charlie being interchanged. Below, we determine the maximum number of sequential Charlies who can be genuinely steered when the three-qubit GHZ state or the W state is  shared between single Alice, single Bob and the sequence of multiple Charlies. 

\subsubsection{When the three-qubit GHZ state is initially shared}\label{sub11}
Let the three-qubit GHZ state (\ref{ghz}) be initially shared between Alice, Bob and multiple Charlies. At first, we  focus on the  following question: how many Charlies, along with the single Bob, can be  genuinely steered by Alice? This is probed through the quantum violation of the inequality (\ref{GHZ1}) as it deals with genuine tripartite ($1 \rightarrow 2$) steering from Alice to Bob-Charlie$^m$. Here, Alice and Bob perform projective measurements, whereas multiple Charlies except the last Charlie perform unsharp measurements. 

We first find out whether Charlie$^1$ and Charlie$^2$ can sequentially violate the inequality (\ref{GHZ1}) with single Bob and single Alice. In this case, the measurements of Charlie$^2$ will be sharp ($\lambda_2 = 1$). We observe that when  Charlie$^1$ gets $G_1 = -0.10$, then Charlie$^2$ gets $G_1 = -0.71$.  This happens for the following choices of measurement settings by Alice:  $( \theta^{x}_0$, $\phi^{x}_0$, $\theta^{x}_1$, $\phi^{x}_1$, $\theta^{x}_2$, $\phi^{x}_2$ $)$ $\equiv$ $( \frac{\pi}{2}$, $0$, $\frac{\pi}{2}$, $\frac{\pi}{2}$, $0$, $0$ $)$, with $\lambda_1 = 0.507$. Here, the choices of measurement settings by Bob, Charlie$^1$ and Charlie$^2$ are not mentioned. Bob and each of the Charlies perform the measurements as specified in the inequality (\ref{GHZ1}). Therefore, the single Bob, Charlie$^1$ and Charlie$^2$ can be genuinely steered  by the single Alice when the GHZ state is initially shared.

Next, we enquire whether Charlie$^1$, Charlie$^2$ and Charlie$^3$ can sequentially violate the inequality (\ref{GHZ1}) with single Alice and single Bob. In this case, the measurements of Charlie$^3$ are sharp ($\lambda_3 = 1$). When Charlie$^1$ gets $G_1 = -0.10$ and Charlie$^2$ gets $G_1 = -0.10$, then Charlie$^3$ gets $G_1 = -0.55$.  This happens for the following choices of measurement settings by Alice: $( \theta^{x}_0$, $\phi^{x}_0$, $\theta^{x}_1$, $\phi^{x}_1$, $\theta^{x}_2$, $\phi^{x}_2$ $)$, $\equiv$ $( \frac{\pi}{2}$, $0$, $\frac{\pi}{2}$, $\frac{\pi}{2}$, $0$, $0$ $)$, with $\lambda_1 = 0.507$ and $\lambda_2 = 0.558$. Hence, Charlie$^1$, Charlie$^2$, Charlie$^3$ and the single Bob can be genuinely steered by the single Alice  when the GHZ state is initially shared.
Proceeding further in this way, we find that at most six sequential Charlies, along with the single Bob, can be genuinely steered by the single Alice.  Here, genuine tripartite EPR steering is probed by the quantum violation of the inequality  (\ref{GHZ1}).

Further, we investigate how many Charlies  can be  genuinely steered by Alice-Bob ($2 \rightarrow 1$ steering) when the three-qubit GHZ state is initially shared between Alice, Bob and Charlie$^1$. In this case, the maximum number of Charlies turns out to be three and this is analyzed using the inequality (\ref{GHZ2}). These results are summarized in Table \ref{tab3}. In this scenario the number of sequential observers as well as the range of sharpness are higher in the ($1 \rightarrow 2$) steering compared to the ($2 \rightarrow 1$) steering cases. 

{\centering
	\begin{table}[t]\footnotesize
			\begin{tabular}{|c | c || c |} 
				\hline 
				& \textbf{Genuine tripartite} & \textbf{Genuine tripartite} \\
				& \textbf{($1 \rightarrow 2$) EPR steering} & \textbf{($2 \rightarrow 1$) EPR steering} \\
				& \textbf{from Alice} & \textbf{from Alice-Bob} \\
				& \textbf{to Bob-Charlie$^m$} & \textbf{to Charlie$^m$}\\
				\hline
				{\bf Charlie$^m$} & {\bf Permissible $\lambda_{m}$ range}  & {\bf Permissible $\lambda_{m}$ range}  \\ [0.5ex] 
				\hline
				\hline
				Charlie$^1$ & 1 $\geq \lambda_1 > \lambda_1^{\text{min}}$ = 0.441  & 1 $\geq \lambda_1 > \lambda_1^{\text{min}}$ = 0.584 \\ 
				\hline
				Charlie$^2$ & 1 $\geq \lambda_2 > \lambda_2^{\text{min}}$ = 0.473 & 1 $\geq \lambda_2 > \lambda_2^{\text{min}}$ = 0.668 \\
				& when $\lambda_i = \lambda_i^{\text{min}}$ $\forall$ $i < 2$ & when $\lambda_i = \lambda_i^{\text{min}}$ $\forall$ $i < 2$ \\
				\hline
				Charlie$^3$ & 1 $\geq \lambda_3 > \lambda_3^{\text{min}}$ = 0.514 & 1 $\geq \lambda_3 > \lambda_3^{\text{min}}$ = 0.805 \\
				& when $\lambda_i = \lambda_i^{\text{min}}$ $\forall$ $i < 3$ & when $\lambda_i = \lambda_i^{\text{min}}$ $\forall$ $i < 3$ \\
				\hline
				Charlie$^4$ & 1 $\geq \lambda_4 > \lambda_4^{\text{min}}$ = 0.568  & No valid range for $\lambda_4$ \\
				& when $\lambda_i = \lambda_i^{\text{min}}$ $\forall$ $i < 4$ & for any $\lambda_i$ with $i<4$ \\
				\hline
				Charlie$^5$ & 1 $\geq \lambda_5 > \lambda_5^{\text{min}}$ = 0.644  & \\ 
				& when $\lambda_i = \lambda_i^{\text{min}}$ $\forall$ $i < 5$ &  \\
				\hline
				Charlie$^6$ & 1 $\geq \lambda_6 > \lambda_6^{\text{min}}$ = 0.763  & \\
				& when $\lambda_i = \lambda_i^{\text{min}}$ $\forall$ $i < 6$ &  \\
				\hline
				Charlie$^7$ & No valid range for $\lambda_7$  & \\ 
				& for any $\lambda_i$ with $i<7$ & \\
				\hline
			\end{tabular}
			\caption{The permissible ranges of the sharpness parameters $\lambda_m$ (where $0 < \lambda_m \leq 1$) of Charlie$^m$ such that they are genuinely steered when the three-qubit GHZ state  is initially shared. }
			\label{tab3}
	\end{table}
} 

\subsubsection{When the three-qubit W state is initially shared}
Finally, we consider the three qubit W state given by Eq.(\ref{wstate}) to be initially shared between single Alice, single Bob and multiple Charlies in both the steering scenarios.

 Following the approach mentioned in Section \ref{sub11}, we observe that at most four Charlies, along with the single Bob, can be genuinely steered by the single Alice in the ($1 \rightarrow 2$) steering scenario. This result is valid in the context of the genuine tripartite EPR steering inequality (\ref{W1}). On the other hand, we  observe that at most three Charlies can be genuinely steered by Alice-Bob for ($2 \rightarrow 1$) steering, which is analyzed through the quantum violation of the inequality (\ref{W2}). These results are summarized in Table \ref{tab4}. Here too, the number of sequential observers as well as the range of sharpness are higher for the ($1 \rightarrow 2$) steering comapred to the ($2 \rightarrow 1$) steering cases.


{\centering
	\begin{table}[t]\footnotesize
			\begin{tabular}{|c | c || c |} 
				\hline 
				& \textbf{Genuine tripartite} & \textbf{Genuine tripartite} \\
				& \textbf{($1 \rightarrow 2$) EPR steering} & \textbf{($2 \rightarrow 1$) EPR steering}\\
				& \textbf{from Alice} & \textbf{from Alice-Bob} \\
				& \textbf{to Bob-Charlie$^m$} & \textbf{to Charlie$^m$} \\
				\hline
				{\bf Charlie$^m$} & {\bf Permissible $\lambda_{m}$ range}  & {\bf Permissible $\lambda_{m}$ range}  \\ [0.5ex] 
				\hline
				\hline
				Charlie$^1$ & 1 $\geq \lambda_1 > \lambda_1^{\text{min}}$ = 0.522  & 1 $\geq \lambda_1 > \lambda_1^{\text{min}}$ = 0.634  \\ 
				\hline
				Charlie$^2$ &1 $\geq \lambda_2 > \lambda_2^{\text{min}}$ = 0.578 &1 $\geq \lambda_2 > \lambda_2^{\text{min}}$ = 0.747 \\
				& when $\lambda_i = \lambda_i^{\text{min}}$ $\forall$ $i < 2$ & when $\lambda_i = \lambda_i^{\text{min}}$ $\forall$ $i < 2$ \\
				\hline
				Charlie$^3$ &1 $\geq \lambda_3 > \lambda_3^{\text{min}}$ = 0.659 &1 $\geq \lambda_3 > \lambda_3^{\text{min}}$ = 0.962 \\
				& when $\lambda_i = \lambda_i^{\text{min}}$ $\forall$ $i < 3$ & when $\lambda_i = \lambda_i^{\text{min}}$ $\forall$ $i < 3$ \\
				\hline
				Charlie$^4$ &1 $\geq \lambda_4 > \lambda_4^{\text{min}}$ = 0.882 &No valid range for $\lambda_4$  \\
				& when $\lambda_i = \lambda_i^{\text{min}}$ $\forall$ $i < 4$ & for any $\lambda_i$ with $i<4$ \\
				\hline
				Charlie$^5$ & No valid range for $\lambda_5$  &  \\
				& for any $\lambda_i$ with $i<5$ &  \\
				\hline
			\end{tabular}
			\caption{The permissible ranges of the sharpness parameters $\lambda_m$ (where $0 < \lambda_m \leq 1$) of Charlie$^m$ such that they are genuinely steered when the three-qubit W state  is initially shared. }
			\label{tab4}
	\end{table}
}

\section{Summary and discussions} \label{s4}

Multipartite quantum correlations are potentially important resources in  quantum networks for accomplishing various quantum communication  tasks. However, due to the difficulties in experimentally producing and preserving  multipartite quantum correlated states from ubiquitous noise, exploring the possibilities of using single copies of multipartite quantum correlation several times is  relevant for  foundational studies. The obstacles in handling multiple copies of  genuine multipartite entangled states in practical scenarios is a strong motivation to investigate how one can partially preserve genuine  EPR steerability of a single copy of a multipartite quantum state even after performing a few rounds of local operations by multiple observers. 

In the present study, we take an initial step in this direction by analyzing thoroughly the scenario of genuine tripartite steering with a sequence of multiple observers. We have considered  three spatially separated spin-$\frac{1}{2}$ particles with multiple observers performing measurements  on one of the particles sequentially and independently of each other, while two  observers perform measurements on  the other two particles in order to implement either ($1 \rightarrow 2$) or ($2 \rightarrow 1$) genuine tripartite steering. We have demonstrated that it is indeed possible for multiple observers to sequentially implement genuine tripartite EPR steering when a three-qubit GHZ state or a W state is initially shared. The results presented here take us a step closer towards performing sequential quantum communication tasks in hybrid quantum networks by utilizing the correlation in a single copy of a multipartite entangled state.

We have obtained the upper bounds on the number of observers who can implement  ($1 \rightarrow 2$) or ($2 \rightarrow 1$) genuine tripartite EPR steering for the above states. The GHZ state turns out to be more powerful compared to the W state allowing for a higher number of observers. Moreover, ($1\rightarrow 2$) steering is more efficient in terms of a larger range of allowed sharpness parameter values compared to that for the ($2 \rightarrow 1$) steering cases. In an earlier work \cite{Maity2} it was shown that the number of observers who can sequentially detect genuine tripartite entanglement for the three-qubit GHZ state is  twelve, and that for the three-qubit W state is four. On the other hand,  at most two sequential observers can demonstrate genuine tripartite nonlocality with the other two spatially separated observers  when the three-qubit GHZ state is initially shared, while only one observer can do so when the shared state is the three-qubit W state \cite{Saha}.  Our present paper complements the above studies \cite{Saha, Maity2} on the analyses of preserving the three categories of genuine tripartite quantum correlations, {\it viz.} entanglement, steerabilty and Bell nonlocality,  after performing a few rounds of measurements on a single quantum state.  In particular, genuine tripartite entanglement is necessary for demonstrating genuine tripartite EPR steering, and genuine tripartite steering is necessary for genuine tripartite nonlocality. Hence, the present study, together with the earlier two studies \cite{Saha, Maity2},  reveals how the maximum number of sequential observers who can detect  quantum correlation, differs in the context of the aforementioned three inequivalent forms of genuine tripartite quantum correlations. 

 Applications of the present study can  be demonstrated in quantum secret sharing protocols. Recently, multipartite genuine EPR steering has been shown to be related with quantum secret sharing protocols \cite{Hil}. Secret sharing is a cryptography protocol which is relevant for practical and commercial quantum communication. In this protocol, a dealer (say, Alice)  sends a message to players (say, Bob and Charlie) in  such a way that the message can be decoded only if Bob and Charlie collaborate to act together. The efficacy of this protocol is linked with the concept of tripartite steering \cite{Daniel,Armstrong2015,sss1,sss2}. In this context, our results point out that Alice can share a secret message with a single Bob and multiple sequential Charlies as well. However, one has to compensate with a reduced quantum advantage in this scenario since multiple Charlies have to perform unsharp measurements in order to sequentially demonstrate tripartite steering. This direction thus illustrates the potential of our results in sharing secret messages with a larger number of players using only one copy of a tripartite steerable state.  Determining the precise relationship between the security of this sequential quantum secret sharing protocol and sequential detection of genuine tripartite EPR steering merits further investigation.

The present results have implications on the security of cryptography protocols where genuine tripartite EPR steering is necessary. Suppose a tripartite genuinely  EPR steerable state is prepared and shared between three parties, say, Alice, Bob and Charlie. While passing one particle to Charlie from the source, it may be intercepted by an eavesdropper. It is evident from our analysis that tripartite EPR steering can still be detected by Charlie even when the eavesdropper disturbs the state by performing up to a certain number of local unsharp measurements, and then passes the particle to Charlie. Hence, this interjection by the eavesdropper may not be noticed if steerability is tested as the criterion for security post the above-mentioned eavesdropping.

Before concluding, it may be noted that experimental verification of our results, being valid in the intermediate scenario between  genuine tripartite entanglement detection and genuine tripartite nonlocality, should be feasible since steerability is more tolerant to environmental noise than the sequential genuine nonlocality detection case \cite{Saha}, and has less difficulties in experimental realization than the sequential genuine entanglement witnesses \cite{Maity2}. As sequential sharing of two-qubit nonlocality \cite{exp1,exp2,expnew}, sequential quantum random access code \cite{exp4,rac1} and sequential sharing of two-qubit  steering \cite{Choi2020} have already been experimentally demonstrated, our analysis is amenable for experimental verification  in the near future. Finally, the analysis presented here could be extended by employing more general unsharp measurement formalisms \cite{colbeck},  with the aim of further increasing the number of parties who can demonstrate genuine tripartite steering through a single copy of the
prepared quantum state.

\section{ ACKNOWLEDGEMENTS}
DD acknowledges Science and Engineering Research Board (SERB), Government of India for financial support through National Post Doctoral Fellowship (NPDF). ASM acknowledges support from the DST project DST/ICPS/QuEST/2019/Q98.


\end{document}